\documentstyle[prd,aps,preprint,amsfonts]{revtex}

\begin{document}

\draft

\title{Aharonov-Bohm Scattering of a Localized Wave Packet:\\
Analysis of the Forward
Direction\thanks
{This work is supported in part by funds provided by the U.S.
Department of Energy (D.O.E.) under cooperative
agreement \#DF-FC02-94ER40818.}}

\author{D. Stelitano}

\address{Center for Theoretical Physics \\
Laboratory for Nuclear Science \\
and Department of Physics \\
Massachusetts Institute of Technology \\
Cambridge, Massachusetts 02139\\
e-mail:stelitan@mit.edu}

\date{MIT-CTP\#2383  \qquad \qquad hep-th/9411175\\
submitted to {\em Physical Review D}}

\maketitle

\begin{abstract}
The Aharonov-Bohm scattering of a localized wave packet is
considered. A careful analysis of the forward direction points out new
results: according to the time-dependent solution obtained by means of
the asymptotic representation for the propagator (kernel), a
phenomenon of auto-interference occurs along the forward direction,
where, also, the probability density current is evaluated and found
finite.
\end{abstract}


\narrowtext

\section{Introduction}

In 1949 W.~Eherenberg and R.~E.~Siday \cite{ES} discussed the effects
of electromagnetic vector potentials on the phases of quantum
mechanical wave functions. Ten years later, Y.~Aharonov and D.~Bohm
\cite{AB} analyzed, from a theoretical point of view, the physical
consequences of the introduction in the empty space of an infinite
solenoid whose radius tends asymptotically to zero while the total
flux of the magnetic field inside it is kept constant.  One of the
results of this work was the anticipation of the phenomenon of
interference (later called Aharonov-Bohm effect) that occurs when two
coherent electron beams pass by each side of a very long, very tiny
solenoid and are subsequently recollected.  Furthermore, the AB
scattering of a plane wave by such a flux tube (a vortex-like magnetic
potential) was studied in the same work and the scattering amplitude
found for every value of the scattering angle $\vartheta$ except the
forward direction where it seemed to diverge.  In the following years,
the Aharonov-Bohm problem became very popular because of its physical
and philosophical implications on the (discussed) reality of the
electromagnetic potentials \cite{BoSi}.  Nowadays, the AB interaction
is re-encountered in other fields of physics: its planar dynamics
reappears in the Chern-Simons electrodynamics of two interacting
point-like particles \cite{Ja}, and in the quantized Hall effect
\cite{PrGi}.  In $(2+1)$-dimensional gravity, the scattering of a
particle on a cone (closely related to the scattering by infinite
cosmic strings) has been shown to be formally equivalent to the
particle-vortex system \cite{JaDe}.

In spite of the fact that the time-independent quantum mechanical
scattering theory for the AB problem has been investigated by many
different approaches, only very few works on the time-dependent
scattering appear in literature \cite{Kr1}.  Here we carry out the
analysis of what happens to a localized wave packet with generic
impact parameter when it is scattered by a magnetic vortex.  As a new
result, we show that there are no discontinuities in the
time-dependent wave function and that an ``auto-interference'' effect
occurs in the forward direction. Owing to this phenomenon, the probability
density current of the transmitted wave packet varies dramatically
according to the values of the magnetic flux; for example, in case of
odd half-integer values it practically vanishes (for the stationary
case see \cite{AB}).

It thus seems that the AB problem does not
satisfy the usual definition of ``scattering'' so that, as it will be
later shown, a plane wave description of the incident flux no longer
succeeds in describing correctly the physics in the forward direction.

In quantum mechanical Schr\"{o}dinger theory, the time-evolution of the
wave function $\psi$, representing the physical system, is governed by
\begin{equation}
i\hbar \frac{\partial}{\partial t}\, \psi
\left(\bbox{r}; t \right)
           = H \, \psi ( \bbox{r}; t)
\text{ ,}
\label{Schr}
\end{equation}
the Hamiltonian $H$ in the presence of a magnetic field
$\bbox{B}(\bbox{r})=
{\bbox \nabla}\times\bbox{A}$ is
\begin{equation}
H=\frac{1}{2m}
           \left (\bbox{p} -
                      \frac{e}{c}\, \bbox{A}
                      (\bbox{r})\right )^{2}
\text{ .}
\label{Hami}
\end{equation}

Thanks to the cylindrical symmetry of the AB set-up, one can reduce
the analysis to a two dimensional problem, disregarding the
coordinative along the solenoid axis.  If $\Phi$ is the flux of the
magnetic field $B(\bbox{r})$
\begin{equation}\Phi =
\int{d^{2}\bbox{r}\,B(\bbox{r})}\text{ ,}\end{equation}
then the particle-vortex interaction is described by means of the
following vector potential, diverging at the origin of the XY-plane,
\begin {equation}
\bbox{A}(\bbox{r})
           =-\frac{\Phi}{2\pi r^{2}}
		\left( y \, {\bf \hat {\hbox {\boldmath$\imath$}}}
                      -x \,\, {\bf \hat {\bbox {\!\jmath}}} \right)
           =\frac{\Phi}{2\pi}\,{\bbox {\nabla}}\vartheta
           =\frac{\hbar c}{e}\,\nu \bbox{\nabla} \vartheta
\label{vpot}
\end {equation}
with
\begin{equation}
\nu\equiv\frac{e\Phi}{2\pi\hbar c}
\text{ ,}
\end{equation}
and the magnetic field is expressed by a delta-function
\begin{equation}
B(\bbox{r})
           = \bbox{\nabla}\times\bbox{A}
           =\Phi\,\hbox{\large $\delta$}^{2}(\bbox{r})
\text{ .}
\end{equation}

The solution of (\ref{Schr}) with a certain initial condition is
formally given in terms of the Aharonov-Bohm propagator $K$
\begin{equation}
\psi(\bbox{r},t)
           =\int{d^{2}\bbox{r}^{\prime} \,
           \psi(\bbox{r}^{\prime},0)\,
           K\left(\bbox{r}, \bbox{r}^{\prime};t;\nu\right)}
\text{ ,}
\end{equation}
the kernel $K$ can be evaluated using a basis ${\cal B}$ in the
Hilbert space ${\cal H}$ of the wave functions; for example
\begin{eqnarray}
{\cal B} &=& \left\{u^{n}_{E}(r)\,\frac{e^{in\vartheta}}{\sqrt{2\pi}}
           \, e^{-iEt/\hbar} \right\}
           \nonumber\\
u^{n}_{E}(r)&=& J_{\left|n-\nu\right|}(kr)
           \qquad
           \left(k^{2}=\frac{2mE}{\hbar^{2}}\right)
\text{ ,}
\end{eqnarray}
where $J_{\alpha}$ is the Bessel function of the first kind of order
$\alpha$.  Then, setting the initial time $t_{\text{in}}=0$,
\footnote{We remark here that the propagator and, consequently, also
$\psi(\bbox{r},t)$ depend upon the choice of the
particular gauge for $\bbox{A}(\bbox{r})$ only
through a phase factor.  For a suitable choice of the gauge it is also
possible to make the vector potential ``disappear'' altogether from
(\ref{Hami}); in this case, however, the non-trivial boundary
conditions that must be satisfied by functions in ${\cal H}$ prevent
(\ref{Schr}) from being the equation describing the non-interacting
case.  For the sake of simplicity, we shall choose the gauge in which
$\bbox{A}(\bbox{r})$ is expressed by
(\ref{vpot}) and the angular momentum operator (z-component) by
$-i\hbar \partial/\partial \vartheta$; this implies that the kernel
$K$ and the wave function $\psi(\bbox{r},t)$ are
single-valued (see, for example, \cite{Kr2}).}
\begin{eqnarray}
&K&\left(\bbox{r}, \bbox{r}^{\prime};t;\nu \right)
           = \frac{1}{2\pi}   \sum_{n=-\infty}^{\infty}
           \int ^{\infty}_{0} kdk
           \nonumber\\
&&         \quad \times \,  e^  {-i \hbar k^{2} t/2m}
           e^{in(\vartheta-\vartheta^{\prime})}
           J_{|n-\nu|}(kr) \,  J_{|n-\nu|}(kr^{\prime})
\text{ .}
\end{eqnarray}

After performing the integration in the linear momentum $k$
and replacing $n$ by $-n$, the
propagator appears as a sum over the angular momentum eigenvalues
\begin{eqnarray}
&K&\left(\bbox{r},\bbox{r}^{\prime};t;\nu \right)
           = \frac{m}{2\pi i\hbar t} \,
           e^{i\frac{m}{2\hbar t}(r^{2}+r^{\prime 2})}
           \nonumber\\
&&         \quad \times   \sum_{n=-\infty}^{\infty}
           {e^{-in(\vartheta-\vartheta^{\prime})}
           e^{-i|n+\nu|\pi/2}
           J_{|n+\nu|}({\scriptstyle\frac{mrr^{\prime}}{\hbar t}})}
\text{ .}
\label{propag}
\end{eqnarray}

It is also possible to evaluate the propagator using path-integral
techniques over multiply-connected spaces (see, for example, \cite{Sh}).

\section{The Initial Wave Packet}\label{Inipa}

We shall first suppose that the particle at the initial time $t= 0$ is
described by a (normalized) Gaussian wave packet centered in
$(-r_{\!o},0)$, with an average wave vector $(k_{o},0)$, that is with
a vanishing impact parameter ($b=0$)
\begin{eqnarray}
\psi_{\text{gauss}}(\bbox{r},0)
&=&
           \frac{1}{\sqrt{2\pi} \, \xi}
           e^{i \bbox{k}_{o}\cdot\bbox{r}
           -{\left(\bbox{r}
           -\bbox{r}_{o}\right)^{2}}/
           {4\xi^{2}}}
           \nonumber\\
&=&        \frac{1}{\sqrt{2\pi} \, \xi}\,
           e^{ik_{o}r\cos\vartheta -
           \left(r^{2}+r_{o}^{2}+2rr_{o}\cos\vartheta\right)/4\xi^{2}}
\end{eqnarray}
then, the propagating wave function is given by
\begin{eqnarray}
\psi(\bbox{r},t)
    &=&    \int d^{2}\bbox{r}^{\prime}
           { \frac{1}{\sqrt{2\pi} \, \xi}}\,
           e^{i\bbox{k}_{o}
           \cdot \bbox{r}^{\prime}
           -{\left(\bbox{r}^{\prime}
           -\bbox{r}_{o}\right)^{2}}/{4\xi^{2}}}\,
           K\left(\bbox{r},
           \bbox{r}^{\prime};t;\nu\right)
           \nonumber \\
    &=&    \int \! \!\!\int r^{\prime}dr^{\prime}d\vartheta^{\prime}
           \nonumber\\
    &&     {} \times { \frac{1}{\sqrt{2\pi}  \,  \xi}} \,
           e^{ik_{o}r^{\prime} \cos \vartheta^{\prime}
           -\left(r^{\prime  2} + r_{o}^{2}
           + 2r^{\prime}r_{o}\cos\vartheta^{\prime}\right)/4\xi^{2}}
           \nonumber\\
    &&     \quad {} \times  K(r,\vartheta,r^{\prime},
           \vartheta^{\prime};t;\nu)
\text{ .}
\label{wavefun}
\end{eqnarray}

A rough estimate for the spatial and kinematical extensions of the
Gaussian packet is given by the corresponding variances
\begin{mathletters}
\begin{eqnarray}
&\Delta& x
           =  \sqrt{\langle x^{2}\rangle-\langle x\rangle^{2}}
           =\Delta y=\xi
           \text{ ,}\\
&\Delta& k_{x}
           =\sqrt{\langle k_{x}^{2} \rangle - \langle k_{x}\rangle^{2}}
           =\Delta k_{y}
           =\frac{1}{2\xi}
\text{ ,}
\end{eqnarray}
\end{mathletters}
and the condition of minimal uncertainty holds
\begin{equation}
\Delta k_{x}\Delta x=\Delta k_{y}\Delta y=1/2
\text{ .}
\end{equation}

In a typical experimental set-up for non relativistic electrons, we
can assume that the particle is very well localized around
$(-r_{\!o},0)$ at a macroscopical distance from the scatterer
\begin{eqnarray*}
&r_{o}&    \sim 10  \text{ cm} \text{ ,}
           \qquad \xi\sim 10^{-3} \text{ cm}\text{ ,}
           \\
&m&_{e^{-}} \simeq 0.5 \text{ MeV} \text{ ,}
           \qquad E_{\text{cin}}\gtrsim 100  \text{ eV}\text{ ,}
\end{eqnarray*}
so that
\begin{equation}
k_{o} = \sqrt{2mE}/\hbar
\sim 5\cdot 10^{8} \text{ cm}^{-1}
\text{ ,}
\end{equation}
and
\begin{equation}
\Delta k= 1/2\xi \sim 10^{3} \text{ cm}^{-1} \ll k_{o}
\text{ ,}
\end{equation}
therefore
\begin{equation}
k_{o}r_{o}\sim 5\cdot 10^{9}\gg 1
\text{ .}
\end{equation}

The request that the spreading of the wave function is negligible
during the whole experiment is translated into \cite{Mes}
\begin{equation}
\sqrt{\frac{D}{k_{o}}} <  \xi
\text{ ,}
\end{equation}
where $D$ is the total length of the particle flight from its source
to the detector.

In the present experiment $D\gtrsim 2 r_{o}$; when $E_{\text{cin}}\gtrsim
100\text{ eV}$, the condition of negligible spreading is thus seen to
be largely fulfilled.  In this situation, the wave packet remains
localized along the direction of motion and all together propagates
with group velocity $\hbar k_{o}/m$. After it passes the origin, its
peak will be found at a distance $r$ from the scatterer when
\begin{equation}
t  \simeq  mr_{o} / \hbar k_{o} + mr /  \hbar k_{o}
\text{ ,}
\end{equation}
therefore, we shall assume that
\begin{equation}
k_{o}/2  <  mr  / \hbar t < k_{o}
\text{ .}
\end{equation}

Finally, simple arguments suggest us to look at the integral
(\ref{wavefun}) as if the integration were in fact restricted to
values of $\bbox{r}^{\prime}$ such that ($\Delta r_{o}$ and $\varepsilon$
are both positive)
\begin{equation}
\bbox{r}^{\prime}  \in [r_{\!o}-\Delta r_{o},r_{\!o}
           +\Delta r_{o}]\otimes[\pi-\varepsilon,\pi+\varepsilon]
\end{equation}
with $\Delta r_{o}$ the same order of magnitude as $\xi$ and $\varepsilon$
as $\xi/r_{\!o}$. Outside these limits the Gaussian function causes
the integrand to vanish.  As a consequence
\begin{equation}
\frac{mrr^{\prime}}{\hbar t} \sim k_{o}r_{\!o} \sim 10^{9} \gg 1
\text{ ,}
\end{equation}
hence we shall replace the exact expression of the propagator by its
asymptotic approximation in the limit $\frac{mrr^{\prime}}{\hbar
t}\rightarrow \infty$.

\section{Asymptotic Expression of the Propagator}

We shall recast the propagator (\ref{propag}) in a way  that
is more convenient for performing the integration (\ref{wavefun})
(see also \cite{AB,Kr1,Sh} for details).  Let
\begin{equation}
\varphi=\vartheta-\vartheta^{\prime}
\qquad
-2\pi<\varphi<2\pi
\text{ ,}
\end{equation}
rescale the summation over the angular momentum eigenvalues, and split
it as follows
\begin{eqnarray}
&K&(r, r^{\prime}, \varphi; t;\nu)
           \nonumber\\
&&         \quad = \frac{m}{2\pi i\hbar t} \,
           e^{i\frac{m}{2\hbar t} (r^{2} + r^{\prime \, 2})}  \,
           e^{i[\nu]\varphi}
           \nonumber\\
&&         \qquad \quad \times \sum_{n=-\infty}^{\infty}
           {e^{-in\varphi}  \,
           e^{-i|n+\{\nu\}|\pi/2}
           J_{|n+\{\nu\}|} ({\scriptstyle \frac{mrr^{\prime}}{\hbar t}}) }
           \nonumber \\
&&         \quad =  \frac{m} {2\pi i\hbar t} \,
           e^{ i \frac{m} {2\hbar t}  (r^{2} + r^{\prime \,  2}) } \,
           e^{i [\nu]\varphi}
           \left(S_{1} + S_{2} + S_{3}\right)
\text{ ,}
\label{rescale}
\end{eqnarray}
where $[\nu]$ is the integer part of $\nu$, $\{\nu\}$ is the
fractional part of $\nu$, so that $\{\nu\}=\nu-[\nu]$ with
$0\leq\{\nu\}<1$ and
\begin{mathletters}
\begin{eqnarray}
S_{1}  &=&  \sum_{n=1}^{\infty}
           e^{-in\varphi} \,
           e^{-i(n+\{\nu\})\pi/2}
           J_{n+\{\nu\}} ({\scriptstyle\frac{mrr^{\prime}}{\hbar t} } )
           \label{tardi} \\
S_{2}  &=&  \sum_{n=1}^{\infty}
           e^{in\varphi} \,
           e^{-i(n-\{\nu\})\pi/2}
           J_{n-\{\nu\}} ({\scriptstyle\frac{mrr^{\prime}}{\hbar t} } )
           \\
S_{3}  &=&  e^{-i\{\nu\}\pi/2}
           J_{\{\nu\}} ({\scriptstyle\frac{mrr^{\prime}}{\hbar t}})
\end{eqnarray}
\end{mathletters}
(note that $S_{2}$ can be obtained by replacing
$\{\nu\}$ by $-\{\nu\}$ and
$\varphi$ by  $-\varphi$ in $S_{1}$).

We have thus separated the propagator into three parts, according to
their positive, negative or zero angular momentum eigenvalues
respectively.  It is possible to turn these sums into integrals,
exploiting appropriate properties of the
Bessel functions \cite{HTF2}. For example, (\ref{tardi}) becomes
\begin{eqnarray}
S_{1} &=&  \frac{1}{2} \,
           e^{-i\{\nu\}\pi/2}
           e^{-i{\scriptstyle\frac{mrr^{\prime}}{\hbar t}} \cos\varphi}
           \nonumber\\
&\times&   \int^{{\scriptstyle\frac{mrr^{\prime}}{\hbar t}}}_{0}
           \!\!  du \, e^{iu\cos\varphi}
           \! \left(J_{\{\nu\}+1}(u)-i
           J_{\{\nu\}}(u)
           e^{\:-i\varphi}\right)
\text{,}
\label{S1}
\end{eqnarray}
where we set the constant of integration to zero because the positive
order Bessel functions in the series vanish at the origin.

Next, split the integration in $S_{1}$ as follows
\begin{equation}
S_{1}={\frac{1}{2}} \, e^{-i\{\nu\}\pi/2} \,
e^{-i{\scriptstyle\frac{mrr^{\prime}}{\hbar t}}\cos\varphi}
\left(I_{1}^{(1)}-I^{(1)}_{2}\right)
\text{ ,}
\end{equation}
with
\begin{eqnarray}
I^{(1)}_{1}  &=&   \int^{\infty}_{0} {\! du \,
           e^{iu\cos\varphi} \!
           \left(J_{\{\nu\}+1}(u)-i
           J_{\{\nu\}}(u)e^{-i\varphi}\right)}
           \text{ ,}\\
I^{(1)}_{2}  &=& \int^{\infty}_{_{{\scriptstyle\frac{mrr^{\prime}}{\hbar t}}}}
           {\!\! du \,
           e^{iu\cos\varphi}  \!
           \left(J_{\{\nu\}+1}(u)-i
           J_{\{\nu\}}(u) e^{-i\varphi}\right)}\text{ ,}
\end{eqnarray}
where the superscript (1) denotes that the expressions refer to
$S_{1}$. Correspondingly
\begin{equation}
S_{2} = {\frac{1}{2}}  \,
           e^{i\{\nu\}\pi/2}   \,
           e^{-i{\scriptstyle\frac{mrr^{\prime}}{\hbar t}}\cos\varphi}
           \left(I_{1}^{(2)}-I^{(2)}_{2}\right)
\text{ .}
\end{equation}

\subsection*{Evaluation of $I_{1}^{(1)}$ and $I_{1}^{(2)}$}

The integral
\begin{equation}
\int^{\infty}_{0} {\!\! du \, e^{iu\cos\varphi}  J_{\alpha}(u)}
\qquad
\Re  (\alpha)  > -1
\end{equation}
is known \cite{TIT1}; however, since $\varphi$ ranges over a
$4\pi$-interval, we have to generalize the solution available for a
$2\pi$-interval. This is simply achieved if one notices that the
solution must be symmetric and periodic in $\varphi$, because so is
the above integral.  Therefore, if $d$ is the branch number over the
log-like Riemann surface (that is the surface obtained by glueing
infinitely many cut donuts), then, almost everywhere,
\begin{equation}
\int^{\infty}_{0} {\!\! du \, e^{iu\cos\varphi}
      J_{\alpha}(u)} = \frac{e^{i \alpha (\pi/2-|\varphi - 2\pi d |)}}
			{|\sin\varphi|}\text{ ,}
\end{equation}
choosing also $-\pi< \varphi - 2\pi d < \pi$ has as a consequence that
$\varphi\rightarrow - \varphi \Rightarrow d \rightarrow -d\/$, so that
periodicity and symmetry are evident.\footnote{The introduction of the
branch number, which causes our results to disagree with those found
in the previous literature, is crucial for the correct description of
the AB problem.}  Hence
\begin{mathletters}
\begin{eqnarray}
I_{1}^{(1)}  &=& \! \frac  {i e^{i\{\nu\}\pi/2}}     {|\sin\varphi|}
           e^{-i \{\nu\} |\varphi - 2\pi d |}
           \left(e^{-i |\varphi - 2\pi d |} - e^{-i \varphi}\right)\!,
           \label{i1}  \\
I_{1}^{(2)}  &=& \! \frac{i e^{ -i \{\nu\} \pi/2}}     {|\sin\varphi|}
           e^{i \{\nu\} |\varphi - 2\pi d |}
           \left(e^{-i |\varphi - 2\pi d |}-e^{ i \varphi} \right)\!.
\label{i2}
\end{eqnarray}
\end{mathletters}

It is easily seen that $I_{1}^{(1)}$ and $I_{1}^{(2)}$ vanish
alternatively on intervals of length $\pi$ and,
for $-2\pi<\varphi<2\pi$, we have
\begin{equation}
{\frac{1}{2}}  \,
           e^{-i\{\nu\}\pi/2}
           I_{1}^{(1)}
           + {\frac{1}{2}} \,
           e^{i\{\nu\}\pi/2}
           I_{1}^{\:(2)}
           =    e^{i \{\nu\} \varphi}
           e^{-2\pi i \{\nu\} d}
\text{ .}\label{i1+i2}
\end{equation}
\subsection*{Evaluation of $I_{2}^{(1)}$ and $I_{2}^{(2)}$}

Supposing $({\scriptstyle \frac{mrr^{\prime}}{\hbar t}})
\longrightarrow \infty$,
we are allowed to replace in $I_{2}^{(1)}$ and $I_{2}^{(2)}$
the integrands with their asymptotic limits.  Using then
\begin{eqnarray}
           J_{\alpha}(u)
&{\:}_{\stackrel {\displaystyle\longrightarrow}
                      {\scriptscriptstyle u\rightarrow\infty }}& \,
           \sqrt{\frac{2}{\pi u}}
	   \cos\left({ u-\alpha \frac{\pi}{2}-\frac{\pi}{4}}\right)
           \nonumber \\
&=&        \frac{1}{\sqrt{2\pi u}}
           \left(e^{i (u-\alpha\frac{\pi}{2} - \frac{\pi}{4})}
           + e^{-i  (u-\alpha \frac{\pi}{2} - \frac{\pi}{4})} \right)
\text{ ,}
\end{eqnarray}
and introducing  the incomplete gamma functions $\Gamma$, we get
\begin{eqnarray}
&I^{(1)}_{2}&
           \nonumber\\
&=&        \!\frac{1}{\sqrt{2\pi }} \,
           e^{-i(\{\nu\}+1)\pi/2}\,
           \frac  {({\scriptstyle 1+e^{-i\,\varphi}})}
           {\sqrt{\scriptstyle 1+\cos\varphi}}\,
           \Gamma[{\scriptstyle \frac {1}{2}}, \,
           -i \, {\scriptstyle\frac{mrr^{\prime}}{\hbar t}}
           (1+\cos\varphi)]
           \nonumber  \\
&+&        \!\!\frac{1}{\sqrt{2\pi }}\,   e^{i(\{\nu\}+1)\pi/2}\,
           \frac {({\scriptstyle 1-e^{-i\,\varphi}})}
           {\sqrt{\scriptstyle 1-\cos\varphi}}\,
           \Gamma[{\scriptstyle \frac {1}{2}},\,
           i \, {\scriptstyle\frac{mrr^{\prime}}{\hbar t} }
          (1-\cos\varphi)]. \!\!
\end{eqnarray}
\widetext

We collect all terms in (\ref{rescale}), expand $S_{3}$ and disregard
two terms that cancel each other for every scattering angle except in
the backward direction, where they are negligible with respect to the
incoming wave packet and therefore can still be neglected.  Finally, the
asymptotic expression of the  propagator is given by
\begin{eqnarray}
&K& \left(r,r^{\prime}, \varphi; t; \nu\right)
           {}_{\stackrel{\displaystyle\Longrightarrow}
           {\scriptscriptstyle \frac{mrr^{\prime}}
		{\hbar t}\rightarrow\infty}}
           e^{i[\nu]\varphi}   e^{i\{\nu\}\varphi}\,
           e^{-2\pi i \{\nu\} d}   \,
           {\frac{m}{2\pi i\hbar t}}      \,
           e^{i\frac{m}{2\hbar t}
           \left(\bbox{r}
           - \bbox{r}^{\prime}\right)^{2}}
           \nonumber \\
&&         {} + {\frac { i} {2\sqrt{2\pi }}} \,
           e^{i[\nu]\varphi}
           \left[  e^{-i\pi \{\nu\}} \,
           \frac {({\scriptstyle 1+e^{-i\varphi}})}
           {\sqrt{\scriptstyle 1+\cos\varphi}}  + e^{i\pi\{\nu\}} \,
	   \frac  {({\scriptstyle 1+e^{i \varphi}})}
           {\sqrt {\scriptstyle 1+\cos\varphi}}\right]
           \nonumber \\
&& \quad   \times \,{\frac{m}{2\pi i\hbar t}}  \,
           e^{i\frac{m}{2\hbar t} (r^2+r^{\prime 2})}
           e^{-i\frac{mrr^{\prime}} {\hbar t} \! \cos\varphi} \;
           \Gamma\left[{\scriptstyle\frac {1}{2}}, -i
           {\scriptstyle \frac{mrr^{\prime}}{\hbar t}}
           (1+\cos\varphi)\right]
           \nonumber   \\
&&         {}+ \sqrt{-i / 2\pi}    \,
           e^{-i\pi\{\nu\}} \,
           e^{i[\nu]\varphi} \,
           {\frac{m}{2\pi i\hbar t}}   \,
           \frac  {e^{i \frac{m}{2\hbar t} (r^2+r^{\prime 2})} }
           {\sqrt{\scriptstyle\frac{mrr^{\prime}}{\hbar t}} }
\text{ .}\label{aspropag}
\end{eqnarray}
As it will be later shown, the first term in (\ref{aspropag}) is
responsible for the transmitted wave packet, the second and
the third one account for the scattered wave function.

\narrowtext
\subsection*{Sommerfeld's integral form of the propagator}

It is possible to transform the sum in (\ref{rescale}) into a contour
integral using the Sommerfeld's integral form of the Bessel function
as given in \cite{HTF2}
\begin{equation}
J_{|n+\{\nu\}|} ({\scriptstyle\frac{mrr^{\prime}} {\hbar t} } )
=  \int_{C_{S}}  \frac{dz}{2\pi} \,
           e^{i {\scriptstyle\frac{mrr^{\prime}} {\hbar t}} \cos{z}}
           e^{i|n+\{\nu\} | (z-\pi/2) }
\text{ ,}
\end{equation}
where $C_{S}$ is a contour from $-\eta+i\infty$ to $2\pi-\eta+i\infty$ with
\begin{equation}
-\eta < 0 < \pi - \eta \qquad
0 < \eta < \pi
\text{ .}
\end{equation}

After some simple manipulations  (see also \cite{Ja}), the kernel is given by
\begin{eqnarray}
K(r, r^{\prime}, \varphi; t;\nu)
&=&   {\frac{m}{2\pi i\hbar t}} \,
           e^{i\frac{m}{2\hbar t}(r^{2} + r^{\prime \, 2})}
           e^{i [\nu]\varphi}
           \nonumber\\
&&         \times \int_{\Sigma}  { {\frac{dz}{2\pi}} \,
           e^{-i{\scriptstyle\frac{mrr^{\prime}} {\hbar t}} \cos{z}}
           \frac{e^{i \{\nu \} z}}  {1-e^{i(z-\varphi)}}  }
\text{ ,}
\end{eqnarray}
where the contour $\Sigma$ is determined by the particular choice of
$\eta$; only for convenience of representation we shall choose
$\eta\rightarrow 0$, so that $\Sigma$ is drawn as in Fig.\ \ref{figu1} or
equivalently in Fig.\ \ref{figu2}.
In the latter we notice two contributions: one coming from the
integration over the loop, and another one resulting by the two
straight lines.  In evaluating the integrals one should keep in mind
that $-2\pi<\varphi<2\pi$ and the integrand is periodic in $\varphi$;
or, in other words, that the poles occur $2\pi$ far from each other.
This means that, using the residue theorem and introducing the branch
number for $\varphi$,\footnote{We disregard here the values
$\varphi=\pm \pi$ because they are a set of measure zero, hence
negligible in the integration (\ref{wavefun}).}
\begin{eqnarray}
&K&(r, r^{\prime}, \varphi; t;\nu)
           \nonumber\\
&&         ={\frac{m}{2\pi i\hbar t}} \,
           e^{i\frac{m}{2\hbar t}
           (r^{2}+r^{\prime 2})}
           e^{i \nu\varphi}
           e^{-2\pi i \{\nu\} d}
           e^{-i{\scriptstyle\frac{mrr^{\prime}}
           {\hbar t}} \cos{\varphi}}
           \nonumber \\
&&\quad    +\, {\frac{m}{2\pi i\hbar t}} \,
           e^{i\frac{m}{2\hbar t} (r^{2}+r^{\prime 2})}
           e^{i[\nu]\varphi}
           \nonumber \\
&&\qquad   \times\int_{SL}  { {\frac{dz}{2\pi}} \,
           e^{-i{\scriptstyle\frac {mrr^{\prime}}{\hbar t}} \cos{z}}
           \frac{e^{i \{ \nu \} z}}  {1-e^{i(z-\varphi)}}  }
\end{eqnarray}
($SL$ refers to the straight line integration).

It is also possible to evaluate the integral over $SL$ for the
scattered wave function in the asymptotic limit
${\frac{mrr^{\prime}}{\hbar t}}\longrightarrow \infty$ and if
$\vartheta$ is out of the forward direction (for the stationary case
see \cite{Ja}): the result is the propagator we shall use in the first
part of the next section.

\section{The Time-Dependent Wave Function}
\label{sec-out}

\subsection{Out of the Forward Direction}

In considering the case\footnote{The factor 2 in
$2\varepsilon$ can of course be replaced by
any reasonable number.}
$2\varepsilon<\vartheta<2\pi-2\varepsilon$
recall, from section \ref{Inipa}, the integral form of the wave function
\begin{eqnarray}
\psi(r,\vartheta,t )     =
&\int\!\! \int& r^{\prime}dr^{\prime} d \vartheta^{\prime}
           \nonumber\\
&\times&   \frac{1}{\sqrt{2\pi} \, \xi} \,
           e^{ik_{o} r^{\prime} \!
           \cos\vartheta^{\prime}
           - \left(r^{\prime \, 2}+r_{o}^{2}+2r^{\prime}r_{o}
           \cos\vartheta^{\prime}\right)/4\xi^{2}}\,
           \nonumber\\
&&         \quad \times   K\left(r,\vartheta,r^{\prime},
           \vartheta^{\prime} ; t; \nu\right)
\text{ ,}
\label{wavefun2}
\end{eqnarray}
and the fact that the integrand is significantly different from zero
only when $|\vartheta^{\prime}-\pi|\leq \varepsilon$. Under these
conditions, the branch number $d$ in (\ref{aspropag}) vanishes and we
can replace the gamma function in the propagator with its
expression for large argument. After some elementary trigonometry
and neglecting the first term in (\ref{aspropag}) since
it produces  a localized wave function propagating (without spreading)
along the forward direction, thus vanishing at these scattering angles,
the kernel out of the forward direction reads
\begin{eqnarray}
&K& \left(r,r^{\prime}, \varphi ; t; \nu\right)
{\:}_{\stackrel{\displaystyle\Longrightarrow}
           {\scriptscriptstyle \frac{mrr^{\prime}}
           {\hbar t}\rightarrow\infty }}
           -  \sqrt{ i / 2\pi} \,
           {\sin{(\pi\{\nu\})}}
           \nonumber\\
&& \qquad  \qquad \times \frac {  e^{i [\nu]\varphi}e^{i \varphi/2}}
           {|\cos{(\varphi/2)}|}  \,
           {\frac{m}{2\pi i\hbar t}} \,
           \frac  {e^{i \frac{m}{2\hbar t}(r^2+r^{\prime  2})} }
           {\sqrt{\scriptstyle\frac{mrr^{\prime}}{\hbar t}} }
\text{ .}
\label{popcorn}
\end{eqnarray}

The evaluation of the scattering wave function as given in
(\ref{wavefun2}) is now easily performed by taking into account
(\ref{popcorn}) and the following
\begin{equation}
\frac{e^{i [\nu]\varphi}  e^{i \varphi/2}}  {|\cos{(\varphi/2)} |}
           \simeq \frac{ -i (-1)^{[\nu]}
           e^{i[\nu]\vartheta}
           e^{i  \vartheta/2}}  {\sin{(\vartheta/2)}}
\text{ .}
\label{salsiccia}
\end{equation}

The result is
\begin{eqnarray}
\psi(\bbox{r},t)    &\simeq&
           {   (-1)}^{[\nu]} \,   {\frac {i  \sqrt{i}}
           {(2\pi)^{3/4}  }  } \, {\sin{(\pi\{\nu\})}}
           \nonumber\\
&\times&   \frac{   e^{i[\nu]\vartheta}   e^{i\vartheta/2}}
           {\sqrt{\xi k_{o}+i  r_{ o } / 2\xi}   \,
           \sin{(\vartheta/2)}\,\sqrt{r}   } \,
           \psi_{\text{(1)free}}  (r,t)\text{ ,}
\end{eqnarray}
where $\psi_{\text{(1)free}} (r,t)$ is the free-propagating
unidimensional Gaussian wave packet such as
$\psi_{\text{(1)free}} (r,{mr_{o}}/{\hbar k_{o}})$
is centered in $r=0$ and has a  positive wave vector $k_{o}$ .
The solution out of the forward direction is
thus seen to be an outgoing circular wave that starts propagating from
the origin at the time $t_{o}\simeq {mr_{o}}/{\hbar k_{o}}$.

Defining $dN_{\text{sc}}/d\vartheta$ and $dN_{\text{inc}}/dL$
respectively as the
number of the particles scattered in a particular direction within a
unitary ``solid'' angle and the number of the particles crossing a
unitary segment perpendicular to the direction of incidence, the
differential scattering cross section is
\begin{equation}
\frac{ d \sigma}{d\vartheta}
           \:  \stackrel{def}{=} \:
           \frac{dN_{\text{sc}}/d\vartheta}   {dN_{\text{inc}}/d\vartheta}
           \:  = \:
           \frac  {\sin^{2}{(\pi\{\nu\})}}
           {2\pi k_{o} \,  \sin^{2} {(\vartheta/2)}}\text{ ,}
\end{equation}
in agreement with the result obtained through time-independent analyses.

\subsection{Forward Direction}

Because of the difficulties one meets when one tries to carry out the
integration (\ref{wavefun2}) with K given by (\ref{aspropag}), the
time-dependent wave function close to the forward direction is not
available to us in an analytical form.  Nevertheless, if one considers
a small neighborhood of $\vartheta=0$, say, for example,
\begin{equation}
\vartheta\in [0,\varepsilon^{2}] \cup [2\pi-\varepsilon^{2}, 2\pi)
\text{ ,}
\end{equation}
it is possible to add some considerations.

In (\ref{aspropag}) the gamma function is of unitary order
\cite{HTF2,Tri}; also, we have
\begin{eqnarray}
e^{i [\nu]\varphi}
\hskip-1.5ex && \hskip-1.5ex
           \left[  e^{-i \pi \{\nu\}}
           \frac  {(1+e^{-i \varphi})}   {\sqrt{1+\cos\varphi}}
           +  e^{i\pi\{\nu\}}
           \frac  {(1+e^{i  \varphi})}   {\sqrt{1+\cos\varphi}}   \right]
           \nonumber \\
&\simeq&
           2\,  (-1)^{[\nu]}  \cos{(\pi\{\nu\})}\,
           \sqrt{1+\cos\vartheta^{\prime}}
           \nonumber \\
&&         \quad - 2 \, (-1)^{[\nu]} \sin{(\pi\{\nu\})}\,
           \frac{\sin\vartheta^{\prime}}
           {\sqrt{1+\cos\vartheta^{\prime}}}
\text{ ,}
\end{eqnarray}
and from suitable arguments about the symmetry
in the angular variable $\vartheta^{\prime}$
one estimates that the contribution to the wave function arising from the
second  term in (\ref{aspropag}) is infinitesimal of order
${\cal O}(\varepsilon)$.  The third term in
(\ref{aspropag}) contributes only through a factor as small as
${\cal O}({\scriptstyle\sqrt{\frac {\hbar t} {mrr^{\prime}} }}) $.
Therefore
\begin{eqnarray}
\psi(r,\vartheta^{\ast},t)
           &&  \nonumber  \\
=\int \!\!\! \int_{0}^{2\pi}
\hskip-1.5ex && \hskip-1.5ex
           r^{\prime} d r^{\prime} d \vartheta^{\prime}
           \frac{1}{ \sqrt{2\pi} \, \xi } \,
           e^{
           i \bbox{k}_{o}
           \cdot \bbox{r}^{\prime}
           -  \left(\bbox{r}^{\prime}
                      - \bbox{r}_{o}\right)^{2} /
                      4\xi^{2} }\,
           \nonumber\\
&&         {} \times e^{i \nu (\vartheta^{\ast} -\vartheta^{\prime}) }\,
           e^{-2\pi i \{\nu\} d}    \,
           \frac{m}{2\pi i \hbar t}  \,
           e^{i\frac{m}{2\hbar t}
                      \left(\bbox{r}^{\ast}
                      - \bbox{r}^{\prime}\right)^{2} }
           \nonumber \\
&& {}+ \left[ {\cal O} ( {\scriptstyle\sqrt {\frac {\hbar t}
		{mrr^{\prime} } } }   )
            +  {\cal O} (\varepsilon) \right]
           \psi_{\text{sc}} (r, \vartheta^{\ast}, t )
\text{ ,}
\label{forwa}
\end{eqnarray}
where the asterisk $\ast$ reminds us that we are very close to the
forward direction.

If we distinguish between the two cases
$ \vartheta\rightarrow 0^{\scriptscriptstyle +}$  and
$\vartheta\rightarrow 2\pi^{\scriptscriptstyle -}$,
we have for the branch number $d$
\begin{enumerate}
\item      $\vartheta \rightarrow 0^{\scriptscriptstyle +}$:
           $$d =  \cases{0&for $0\leq \vartheta^{\prime}<\pi$,\cr
           -1&for $\pi\leq \vartheta^{\prime}<2\pi$.\cr}
           $$
\item      $\vartheta \rightarrow 2\pi^{\scriptscriptstyle -}$:
           $$d =  \cases{1&for $0\leq \vartheta^{\prime}<\pi$,\cr
           0&for $\pi\leq \vartheta^{\prime}<2\pi$.\cr}
           $$
\end{enumerate}
\widetext
Splitting the integration in two parts, the two cases are easily
recognized to be the same one
\begin{eqnarray}
\psi(r,\vartheta^{\ast},t)  &=&
           \int\!\!\!\int_{0}^{\pi}
           r^{\prime}dr^{\prime} d \vartheta^{\prime}
           \frac{1} {\sqrt{2\pi} \, \xi} \,
           e^{ i \bbox{k}_{o} \cdot
                      \bbox{r}^{\prime}
                      -  \left(\bbox{r}^{\prime}
                                 - \bbox{r}_{o} \right)^{2} /
                                   4\xi^{2}  } \,
           e^{-i [\nu]  \vartheta^{\prime} }
           e^{-i \{\nu\} \vartheta^{\prime}}
                      {\frac{m}{2\pi i\hbar t}}      \,
                      e^{i\frac{m}{2\hbar t}
                      \left(\bbox{r}^{\ast}
                      - \bbox{r}^{\prime} \right)^{2} }
           \nonumber  \\
&&       {}+ \int\!\!\!\int_{\pi}^{2\pi} r^{\prime} d
           r^{\prime} d \vartheta^{\prime}
           \frac{1}{\sqrt{2\pi} \, \xi} \,
           e^{ i \bbox{k}_{o}
           \cdot \bbox{r}^{\prime}
           - \left( \bbox{r}^{\prime}
                      - \bbox{r}_{o} \right)^{2} / 4 \xi^{2}  }\,
           e^{-i [\nu]  \vartheta^{\prime} }
           e^{i \{\nu\}(2\pi-\vartheta^{\prime})}
                      \frac{m}{2\pi i\hbar t} \,
           e^{i \frac{m}{2\hbar t}
                      \left( \bbox{r}^{\ast}
                      - \bbox{r}^{\prime}\right)^{2} }
           \nonumber\\
&&      \qquad {}+  \left[ {\cal O}( {\scriptstyle\sqrt{\frac {\hbar t}
		{mrr^{\prime}}    }})
           +  {\cal O} (\varepsilon) \right ]
           \psi_{\text{sc}} (r, \vartheta^{\ast}, t)
\text{ ,}
\label{contifu}
\end{eqnarray}
and the time-dependent wave function is thus seen to be continuous and
single-valued also in the forward direction.

\narrowtext

Next, from  section \ref{Inipa}, we shall replace
$e^{-i [\nu]  \vartheta^{\prime} } e^{-i \{\nu\} \vartheta^{\prime}}$
with $(-1)^{[\nu]}e^{-i \{\nu\}\pi}$ in (\ref{contifu}).
Since for these values of $\vartheta$, the integrands now depend on
$\vartheta^{\prime}$ only through cosine (apart from
a ${\cal O}(\varepsilon)$ factor),
we can rewrite the wave function in a neighborhood of
$\vartheta=0$ (here denoted by $0^{\pm}$) in the following simple form
\begin{eqnarray}
\psi(r,0^{\pm},t) &=&
           (-1)^{[\nu]}  \cos{(\pi\{\nu\})}
           \psi_{\text{free}} (r,0^{\pm},t)
           \nonumber\\
&+&        {\cal O}(\varepsilon)
           \psi_{\text{free}} (r,0^{\pm},t)
           \nonumber\\
&+&        \left[ {\cal O}({\scriptstyle\sqrt{\frac{\hbar t}
		{mrr^{\prime}}}})
           +{\cal O}(\varepsilon)  \right]
           \psi_{\text{sc}}(r,0^{\pm},t)
\text{ ,}
\label{result}
\end{eqnarray}
where $\psi_{\text{free}}$ is the free-propagating bidimensional Gaussian
wave packet.

The unusual physics of the Aharonov-Bohm problem is summarized in the term
\begin{eqnarray}
(&-&1)^{[\nu]}  \cos{(\pi\{\nu\})} \, \psi_{\text{free}} (r,0^{\pm},t)
           \nonumber\\
&&         = \frac{1}{2} \,
           e^{i\pi \nu}\, \psi_{\text{free}} (r,0^{\pm},t)
           +  \frac{1}{2} \,
           e^{-i\pi \nu}\, \psi_{\text{free}} (r,0^{\pm},t)
\text{ .}\label{unuterm}
\end{eqnarray}
We notice that
\begin{itemize}

\item[$\bullet$] if $\{\nu\}=0$ the AB effect is absent.

\item[$\bullet$] if $\{\nu\}=1/2$
           the wave function practically vanishes along the ray
           $\vartheta = 0$.\footnote{
	   The particular case
           $\{\nu\}=1/2$ is actually easier to handle, since one can
           integrate (\ref{S1}) by directly introducing in it the
           analytical forms for the Bessel functions
           involved.}

[The interpretation of this phenomenon cannot be reduced to that one
of ordinary scattering, in which the probability density current of
the transmitted wave packet is only slightly diminuished (fact that is
quantitatively expressed by the well known ``optical theorem'').]

\item[$\bullet$] in the general case the amplitude of the transmitted
wave packet is modulated by a cosine function of the fractional part
$\{\nu\}$ of the magnetic flux.

\end{itemize}

Our interpretation of this phenomenon is that the two components of
the propagator with different signs of the angular momentum
eigenvalues ``prefer'' to pass by either side of the magnetic vortex
or, in other words, to go either up or down along the log-like Riemann
surface [see (\ref{i1})--(\ref{i1+i2})].  They recombine in the
forward direction and they interfere or better ``auto-interfere'' [see
(\ref{forwa})--(\ref{unuterm})] since the two ``splitted'' parts
originate from a unique initial wave function: they have simply
experienced different histories.

Finally, from (\ref{result}), omitting the infinitesimal terms, we
have a simple expression for the current in a neighborhood of
$\vartheta=0$
\begin{equation}
\bbox{J} (r,0^{\pm},t)
           = \cos^{2}{(\pi\{\nu\})} \,
           \bbox{J}_{\text{free}} (r,0^{\pm},t)
\text{ ,}
\end{equation}
where $\bbox{J}_{\text{free}} (r,0^{\pm},t)$ is the current in the
non-interacting case.

The above results are straightforwardly
generalized to initial wave packets with any impact parameters.

\section{Generic Impact Parameter}

Let $b$ be the impact parameter, then the wave packet at the initial time is
\begin{eqnarray}
&\psi&_{\text{gauss}}(r,\vartheta,0)
           =   \frac{1}{\sqrt{2\pi} \, \xi}\,
           e^{i\bbox{k}_{o} \cdot \bbox{r}
           - \left(\bbox{r}
           - \bbox{\rho}\right)^{2} /
           4\xi^{2}}
           \nonumber\\
&&         =\frac{1}{\sqrt{2\pi} \, \xi}\,
           e^{ i k_{o}r \cos\vartheta
           - \left(r^{2} + \rho^{2}
           -2 \rho r\cos (\vartheta
           - \vartheta_{o})\right) / 4\xi^{2} }
\text{ ,}
\end{eqnarray}
where $\bbox{k}_{o} = (k_{o},0) =  k_{o} e^{i0}$, and
\begin{eqnarray}
&\bbox{\rho}&= (-r_{o},b)
           = \sqrt{r_{o}^{2}+b^{2}}\,  e^{i\vartheta_{o} }
	   = \rho e^{i\vartheta_{o}}
           \nonumber\\
&\vartheta_{0}&= \arctan{(-b/r_{0})}
\text{ ,}
\end{eqnarray}
so that the time-dependent wave function is
\begin{eqnarray}
&\psi&     (r,\vartheta,t) = \int\!\!\!\int r^{\prime}
           d r^{\prime} d \vartheta^{\prime}
           \nonumber\\
&&         \times\,  \frac{1} {\sqrt{2\pi} \, \xi}  \,
           e^{ik_{o} r^{\prime} \! \cos \vartheta^{\prime}
           - \left(r^{\prime  2} + \rho^{2} - 2\rho r^{\prime}
           \cos (\vartheta^{\prime}-\vartheta_{o})  \right)
           /4\xi^{2}  }
           \nonumber\\
&&         \times\, K \left (r,\vartheta,r^{\prime},
           \vartheta^{\prime}; t; \nu\right)
\text{ .}
\label{gigetto}
\end{eqnarray}

\subsection{Scattered Wave Function}

Using the following
\begin{equation}
e^{\rho    r^{\prime} \cos(\vartheta^{\prime} - \vartheta_{o})
           / {2\xi^{2}} }
           = \sum_{n = -\infty}^{\infty}
           {e^{in(\vartheta^{\prime} - \vartheta_{o}+\pi/2)}
           J_{n}
           \left(-i\rho r^{\prime} / 2\xi^{2}\right)}\text{ ,}
\end{equation}
the Graaf's addition formula \cite{HTF2} and the definition of $\rho$,
we have that
\begin{eqnarray}
&\int&     d \vartheta^{\prime}   \,
           e^{ ik_{o} r^{\prime} \! \cos \vartheta^{\prime}+\rho
           r^{\prime} \! \cos(\vartheta^{\prime}
           - \vartheta_{o}) / 2 \xi^{2} }
           \nonumber \\
&&         = 2 \pi \! \sum_{n=-\infty}^{\infty}
           {e^{in(\vartheta_{o}+\pi)}
           J_{n}
           \left(k_{o}r^{\prime}\right)
           J_{n} \left(-i \rho r^{\prime} /  2\xi^{2}\right) }
           \nonumber \\
&&         = 2\pi J_{o}
           \left(r^{\prime} {\sqrt{(k_{o}+i{r_{o}} / {2\xi^{2}})^{2}
           - {b^{2}} / {4\xi^{4}}}} \right)
\text{ ,}
\label{Graaf}
\end{eqnarray}
furthermore,  for any reasonable value of $b$
\begin{eqnarray}
&&{\sqrt {\left(k_{o}+i r_{o}/2\xi^{2} \right)^{2}
           - {b^{2}}/{4\xi^{4}}}}
           \nonumber  \\
&&  \qquad \simeq
           \left(k_{o}+i  r_{o}/ 2\xi^{2} \right)
           \sqrt{1- b^{2}/ 4\xi^{4}k_{o}^{2}}
           \nonumber \\
&&  \qquad \simeq
           k_{o} \left({ 1- b^{2}/8\xi^{4}k_{o}^{2}}\right)
           + i  r_{o} / 2\xi^{2}
           \left(1- b^{2} / 8\xi^{4}k_{o}^{2}\right).
\label{eq28}
\end{eqnarray}

After introducing the respective counterparts of
(\ref{popcorn}) and (\ref{salsiccia}) into
(\ref{gigetto}), we shall make use of (\ref{Graaf}) (with the Bessel
function replaced by its asymptotic form) and of (\ref{eq28}).
When
$\left| | \vartheta - \vartheta_{o}| - \pi \right| \geq 2\varepsilon$, then
the scattered wave function corresponding to the initial wave packet with
impact parameter $b$ reads
\begin{eqnarray}
&\psi&_{\text{sc}}(\bbox{r},t) \simeq
           -{\frac  {\sqrt{i}}{(2\pi)^{3/4}  }  }  \,
           {\sin{(\pi\{\nu\})}}
           \nonumber \\
&&         \times \,\frac{  e^{i[\nu](\vartheta-\vartheta_{o})}
           e^{ i (\vartheta-\vartheta_{o})/2}}
           {\sqrt{\xi k_{o} + i   r_{o}/  2\xi} \,
           \left| \cos{[(\vartheta-\vartheta_{o})/2]}
           \right|\, \sqrt{r}   }
           \int^{0}_{-\infty}  dr^{\prime}
           \nonumber \\
&&         \times \,(2\pi \xi^{2})^{-1/4}  \,
           e^{ik_{o}r^{\prime}
           - (r^{\prime}+r_{o})^{2}/4\xi^{2} } \,
           e^{- b^{2}/4\xi^{2}
           + (r_{o}b^{2}/16\xi^{6}k_{o}^{2}) r^{\prime} }\,
           \nonumber \\
&&         \quad\times \sqrt{\frac{m}{2\pi i\hbar t}}   \,
           e^{i\frac{m}{2\hbar t} (r-r^{\prime})^{2} }\text{ ,}
\end{eqnarray}
but $r^{\prime}$ is kept confined by the Gaussian function to a value
around $\rho$, so that
\begin{eqnarray}
&\psi&_{\text{sc}}(\bbox{r},t)
           \simeq   -  {\frac  {\sqrt{i}}{(2\pi)^{3/4}  }  }    \,
           {\sin{(\pi\{\nu\})}}
           \nonumber \\
&&  \quad  \times \,\frac{  e^{i[\nu](\vartheta-\vartheta_{0})}
           e^{i (\vartheta-\vartheta_{o})/2}}
           {\sqrt{\xi k_{o}+i r_{o} / 2\xi} \,
           \left| \cos{[(\vartheta-\vartheta_{o})/2]}
           \right|\, \sqrt{r}  }
           \nonumber \\
&& \quad   \times \, e^{-b^{2}/4\xi^{2}}
           \int^{0}_{-\infty}
           dr^{\prime}\, (2\pi \xi^{2})^{-1/4}  \,
           e^{ik_{o}   r^{\prime}
           -  \left(r^{\prime}+r_{o}\right)^{2} / 4\xi^{2}  }
           \nonumber \\
&& \qquad   \times \,\sqrt{\frac{m}{2\pi i\hbar t}}  \,
           e^{i \frac{m}{2\hbar t} (r-r^{\prime})^{2} }\text{ ,}
\end{eqnarray}
that is
\begin{eqnarray}
&\psi&_{\text{sc}}(\bbox{r},t)   \simeq
           -   {\frac  {\sqrt{i}}{(2\pi)^{3/4} } } \,
           {\sin{(\pi\{\nu\})}}
           \nonumber \\
&& \quad   \times \,\frac{  e^{i[\nu] (\vartheta-\vartheta_{0})}
           e^{i (\vartheta-\vartheta_{o})/2}}
           {\sqrt{\xi k_{o}+i  r_{o}/  2\xi}   \,
           \left| \cos{[(\vartheta-\vartheta_{o})/2]}
           \right| \,\sqrt{r}  }
           \nonumber \\
&& \qquad   \times\, e^{-{b^{2}} / {4\xi^{2}}}\,
           \psi_{\text{(1)free}}(r,t)
\text{ .}
\end{eqnarray}

{}From the similarity with the case $b=0$ considered above, it is
evident that the relative magnitude of the scattering will be
determined by the factor
$$
e^{-{b^{2}}/{4\xi^{2}}} \text{ ,}
$$
which implies
that the scattering significantly occurs only when $b \lesssim \xi$.

\subsection{Transmitted Wave Packet}

Performing an analysis similar to that one of the previous section, it
is also easy to show that, in the case of negligible spreading, the
branch number $d$ is to be set to zero unless the impact parameter $b$
is close to the spatial variance $\xi$ of the initial wave packet;
that is, the auto-interference  effect  is not noteworthy unless the
``overlap'' between the incoming localized wave packet and the source
of the potential (the origin) is relevant.

Therefore, if $b>\xi$ the initial wave packet can be considered to
propagate freely
(still taking the overall  phase factor that accounts for the
presence of the magnetic vector potential \cite{AB}).

\section{Conclusion}

The time-dependent analysis of the Aharonov-Bohm problem points out
that along the forward direction an auto-interference effect occurs
between the components of the incoming wave packet\footnote {The
above time-dependent analysis can be carried out using a
generic initial wave packet that only satisfies the requirements of
being localized and symmetric with respect to the X-axis.
For concreteness, we took a Gaussian function.} that have positive
and negative eigenvalues of the angular
momentum respectively or, more intuitively, that ``pass'' to the right
or to the left of the solenoid.  One can split the wave function in
these two parts (see also \cite{AB})
\begin{equation}
\psi=\psi_{1}+\psi_{2}
\text{ .}
\end{equation}
The superposition of these two components generates the ``differential
cross section'' when the scattering angle is far from the forward
direction, while in this direction (and in a neighborhood of it) also
the auto-interference effect becomes evident, deeply modifying
the probability density current and the shape of the transmitted wave
packet.  Therefore it seems that the classical definitions of
differential and total cross section are no longer meaningful in the
AB set-up.

On the other hand, the time-dependent wave function is
continuous and single-valued and reduces to the free-propagating wave
packet when its impact parameter is larger than its spatial extension
(that is just the classical result). As a consequence, the
``differential cross section'' does not diverge along the forward
direction.

Furthermore, the time-dependent wave function cannot be reduced to a
superposition of stationary eigenfunctions (multiplied by the
corresponding temporal phase factors) $\psi_{k}(r,\vartheta)$
such as
\begin{equation}
\psi_{k}(r,\vartheta) \sim
           e^{ikr\cos{\vartheta}} + \psi_{\text{sc},k}(r,\vartheta)
\end{equation}
with
\begin{equation}
\psi_{\text{sc},k}(r,\vartheta)
{\:}_{\stackrel{\displaystyle\longrightarrow}
           {\scriptscriptstyle r\rightarrow\infty} }
           f_{k}(\vartheta)\,
           \frac{e^{ikr}}{\sqrt{r}}\text{ ,}
\end{equation}
and not even if phase factors are to be introduced in the ``plane'' wave.

For this reason, a time-independent approach performed by the usual
means of the phase shift analysis does not describe properly
the physics of the AB problem.

Finally, we wish to remark that the introduction of the branch number
in the angular variable, responsible in our discussion for the
single-valuedness of the propagator, has also been considered through
a time-independent approach in a recent work \cite{BaBe}.

\acknowledgements

I thank Professor R.~Jackiw for suggesting this problem and
acquainting me with Reference \cite{Ja}. I am also indebted to D.~Bak,
O.~Bergman and G.~G\'erard for many useful discussions.

\begin{figure}

\caption{Integration contour $\Sigma$.}

\label{figu1}

\end{figure}

\begin{figure}

\caption{Equivalent representation for the integration contour $\Sigma$.
         The integration over the loop is responsible for the transmitted
         wave packet, the integration over the straight lines for the
         scattered wave function.}

\label{figu2}

\end{figure}


\begin{thebibliography}{99}

\bibitem{ES}
W. Eherenberg and R. E. Siday,
Proc.\ Phys.\ Soc.\  (London) B {\bf 62}, 8
(1949).

\bibitem{AB}Y. Aharonov and D. Bohm, Phys.\ Rev.
\ {\bf 115}, 485 (1959).

\bibitem{BoSi}P. Bocchieri, A. Loinger and G. Siragusa,
Lett.\ Nuovo Cimento (1978-1981); S. N. M. Ruijsenaars,
Ann.\ Phys.\ (NY) {\bf 146}, 1 (1983).

\bibitem{Ja}R. Jackiw, Ann.\ Phys.\ (NY) {\bf 201}, 83 (1990).

\bibitem{PrGi}{\em The Quantum Hall Effect}, edited by
R. Prange and S. Girvin (Springer Verlag, Berlin, 1990).

\bibitem{JaDe}S. Deser, R. Jackiw and G. 't Hooft, Ann.\ Phys.\ (N.Y.)
{\bf 152}, 220 (1984); G. 't Hooft, Commun.\ Math.\ Phys.\ {\bf
117}, 685  (1988); S. Deser and R. Jackiw, Commun.\ Math.\ Phys.\
{\bf 118}, 495 (1988).

\bibitem{Kr1}M. Kretzschmar, Zeitschrift f\"{u}r Physik
{\bf 185}, 84 (1965).

\bibitem{Kr2}M. Kretzschmar, Zeitschrift f\"{u}r Physik
{\bf 185}, 73 (1965).



\bibitem{Sh}C. C. Bernido and A. Inomata, J. Math. Phys.\ {\bf 22}, 715
(1981); G. Morandi and E. Menossi, Eur.\ J. Phys.\ {\bf 5}, 49
(1984); A. Y. Shiekh, Ann.\ Phys.\ (NY) {\bf 166}, 299 (1986).

\bibitem{Mes}A. Messiah, {\em Quantum Mechanics} (North-Holland, Amsterdam,
1961--62).

\bibitem{HTF2}{\em Bateman Manuscript Project}, edited by
A. Erdelyi {\em et al.}, Higher Trascendental Functions Vol.~II,
(McGraw-Hill, 1953).


\bibitem{TIT1}\samepage{{\em Bateman Manuscript Project}, edited by
A. Erdelyi {\em et al.}, Tables of Integral Transforms Vol.~I,
(McGraw-Hill, 1953).}

\bibitem{Tri}F. G. Tricomi, {\em Funzioni ipergeometriche confluenti},
(Edizioni Cremonese, Roma, 1954).

\bibitem{BaBe}D. Bak and O. Bergman, Massachusetts Institute of Technology,
Cambridge (MA), MIT preprint CTP \# 2283, 1994 (to be published).



\end{thebibliography}
\end{document}